\documentclass[%
 reprint,
 amsmath,amssymb,
 aps,
]{revtex4-1}

\usepackage{graphicx}
\usepackage{dcolumn}
\usepackage{bm}
\usepackage{xcolor}

\begin{document}

\preprint{ }

\title{Bulk self-assembly of giant, unilamellar vesicles}

\author{James T. Kindt}%
\affiliation{%
 School of Chemistry, Emory University, Atlanta, GA 30322, USA}
\author{Jack W. Szostak}
\thanks{Co-corresponding authors}
\affiliation{Department of Molecular Biology and Center for Computational and Integrative Biology, Massachusetts General Hospital, Boston, MA 02114, USA}
\author{Anna Wang}
\thanks{Co-corresponding authors}
 \affiliation{School of Chemistry, UNSW Sydney, NSW 2052, Australia}
 \affiliation{Department of Molecular Biology and Center for Computational and Integrative Biology, Massachusetts General Hospital, Boston, MA 02114, USA}
 

\begin{abstract}
    The desire to create cell-like models for fundamental science and applications has spurred extensive effort towards creating giant unilamellar vesicles (GUVs). \textcolor{black}{However, a route to selectively self-assemble GUVs in bulk has remained elusive. In bulk solution,} membrane-forming molecules such as phospholipids, single-tailed surfactants, and block copolymers typically self-assemble into multilamellar, onion-like structures. So although self-assembly processes can form nanoscale unilamellar vesicles, scaffolding by droplets or surfaces is required to create GUVs. Here we show that surprisingly, it is possible to bulk self-assemble cell-sized GUVs with almost \textcolor{black}{complete selectivity over other vesicle topologies}. The seemingly paradoxical pair of features that enables this appears to be having very \emph{dynamic} molecules at the nanoscale, that create unusually \emph{rigid} membranes. The resultant self-assembly pathway enables encapsulation of molecules and colloids, and can also generate model primitive cells that can grow and divide.
\end{abstract}

\maketitle

Most cells have a single cell membrane, well-separated from internal organellar membranes. However, when membrane-forming 
amphiphilic molecules self-assemble in water, they typically form 
multilamellar, onion-like structures rather than cell-like unilamellar vesicles. The desire to easily create cell-like models has thus spurred extensive interest and effort towards creating unilamellar vesicles. 
Although nanoscale self-assembled unilamellar vesicles
have been reported~\cite{jung_origins_2001, kaler_spontaneous_1989}, 
scaffolding by droplets~\cite{pautot_production_2003, abkarian_continuous_2011, deng_monodisperse_2016, deshpande_octanol-assisted_2016} or surfaces~\cite{angelova_liposome_1986, koksal_nanotube-mediated_2019, kresse_novel_2016} is required to create \emph{giant} unilamellar 
vesicles (GUVs) and a route to selectively self-assemble such systems in bulk has remained elusive. 

Indeed, the requirements for the bulk-assembly of kinetically-stable giant unilamellar vesicles (GUVs) are seemingly incompatible. First, to obtain \textit{unilamellarity}, the vesicles must be able to remodel into single bilayers during vesicle formation and escape kinetic traps. As a result, instances of bulk-assembled unilamellar vesicles are typically composed of single-tailed surfactants~\cite{jung_origins_2001, kaler_spontaneous_1989}. Single-tailed surfactants, however, form bilayer membranes that have a bending modulus on the order of the thermal energy $kT$, and thus the reported bulk-assembled unilamellar vesicles are nanoscale rather than cell-sized, maximizing entropy~\cite{jung_origins_2001}. \textcolor{black}{Nanoscale unilamellar vesicles can also be prepared by ethanol injection~\cite{batzri_single_1973}}. While vesicles of this size are useful for bulk studies, their utility is restrictive on two fronts: Their imaging requires electron microscopy, which limits the number of timepoints that can be studied during dynamic events such as vesicle division~\cite{berclaz_matrix_2001}, and their small encapsulated interior volume places limitations on vesicle loading. However, molecules that self-assemble into \textcolor{black}{\emph{giant} vesicles are typically also kinetically trapped. Thus, while giant vesicles can form easily~\cite{walde_giant_2010, morigaki_giant_2002, moscho_rapid_1996, bagatolli_giant_2000, laouini_preparation_2012}, the production of giant \emph{unilamellar} vesicles requires remodelling the membranes into a unilamellar form with laser heating~\cite{oana_spontaneous_2009}, electric fields~\cite{angelova_liposome_1986}, surface-adsorption~\cite{koksal_nanotube-mediated_2019, kresse_novel_2016, morigaki_giant_2002, horger_films_2009, lira_giant_2014}, emulsion droplets~\cite{pautot_production_2003, abkarian_continuous_2011}, or droplet-based microfluidics~\cite{deng_monodisperse_2016, deshpande_octanol-assisted_2016}.} The likelihood of bulk GUV assembly thus seems low. 

Here we show that it is possible to \textcolor{black}{\emph{selectively} assemble cell-sized GUVs from fatty acids in bulk solution}, without the need for scaffolding or microfluidics. Counterintuitively, we found that membranes that have a higher charge density and which are therefore more mutually repellant tended to form multilamellar vesicles, whereas membranes with less surface charge generated giant and uniformly unilamellar vesicles. Combining experimental and simulation results, we propose a model for the mechanism of GUV self-assembly based on the protonation-state-dependence of the bending modulus of the membrane. We anticipate that the system can be used in a variety of applications because the GUVs can encapsulate small molecules and even colloidal particles. We show one such use by demonstrating that model protocells consisting of oleate GUVs containing RNA could be triggered to grow and spontaneously divide by the addition of excess oleate in the form of micelles. Our results demonstrate the \textcolor{black}{selective} self-assembly of cell-sized unilamellar vesicles, and point towards surprising, unexplored properties of fatty acids and a potential new model system for origins of life and biophysical studies.

\section{Results and discussion}
\begin{figure}[h]
 \includegraphics[width=0.5\textwidth]{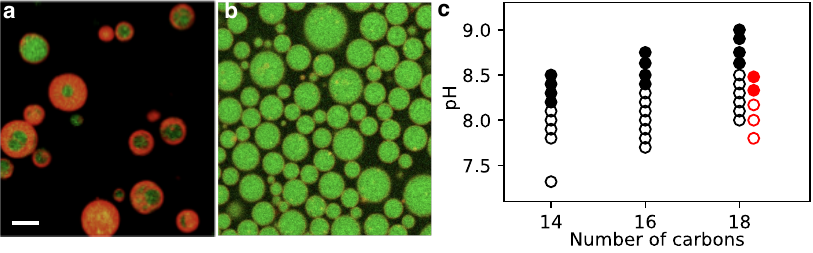}
\caption{Fatty acid self-assembly depends on the aqueous solution conditions. a. Confocal microscopy revealed that oleic acid self-assembled into giant vesicles capable of encapsulating and retaining RNA oligomers (green). The membrane was dyed with 10 $\mu$M rhodamine B (red). The giant vesicles that self-assembled in 200 mM Na$^+$ bicine, pH 8.43 were very heterogeneous in morphology. b. The giant vesicles that self-assembled in 50 mM Na$^+$ bicine, pH 8.43 appeared to be unilamellar. c. The pHs at which GUVs assembled for myristoleic acid (14 carbons), palmitoleic acid (16 carbons) and oleic acid (18 carbons) in 50 mM Na$^+$ bicine are shown as open circles (black), and the pHs at which MLVs formed are shown as solid circles. The red circles show the self-assembly of oleic acid MLVs (solid circles) and GUVs (open circles) in 250 mM Na$^+$ bicine. The scale bar represents 10 $\mu$m.}
\label{fig:1}
\end{figure}

We found that instead of self-assembling into heterogeneous structures (Fig.~\ref{fig:1}a), fatty acids could, under certain conditions, \emph{selectively} self-assemble into giant, apparently unilamellar vesicles. (Fig.~\ref{fig:1}b). Although it is well-known that fatty acids exhibit a rich, pH-dependent variety of self-assembly processes, forming micelles at high pH, neat oil or cubosomes at low pH, and bilayer membranes at a pH near the apparent p$K_a$~\cite{suga_characterization_2016, gebicki_preparation_1976}, we discovered that within the pH range compatible with vesicle formation lies a previously unexplored divergence in self-assembly behaviors. 

We dispersed oleate micelles \textcolor{black}{(pH $>$ 10) into more acidic solutions, such as dilute HCl (Video S1), to yield solutions with a final pH in the range of 8.1 to 8.8. We then} left the samples overnight on an orbital shaker to allow the self-assembly processes to evolve towards steady-state structures~\cite{maeda_hydrogen_1992}. At pH values from 8.1 to 8.4, just below the reported apparent p$K_a$ of oleic acid, optically transparent solutions formed. Under phase contrast microscopy, however, the solutions were revealed to contain abundant cell-sized spherical vesicles (Fig.~S2). By contrast, at final pH values between 8.5 to 8.8, microscopy revealed the presence of very heterogeneous multilamellar vesicles. The same pH-dependence was also found for palmitoleic acid and myristoleic acid vesicles (Fig.~\ref{fig:1}c). The differences between multilamellar samples and less multilamellar ones were readily apparent by eye, and corresponded well with bulk sample turbidity~\cite{wang_core-shell_2019} (Fig.~S3). \textcolor{black}{Because the growth of giant vesicles on hydrocarbon-covered glass slides has been previously reported~\cite{morigaki_giant_2002}, we took great care during cleaning (see Methods) and imaging (Fig.~S4) to distinguish between bulk-grown vesicles and surface-grown ones}. We found that the morphology also depended on salt concentration (\textcolor{black}{Fig.~\ref{fig:1}c and} Fig.~S5), as expected due to the salt-dependence of the p$K_a$ of oleate in the bilayer membrane phase~\cite{maeda_hydrogen_1992}.\textcolor{black}{Indeed, at higher salt concentrations, the self-assembly of GUVs was only possible at slightly lower pHs (Fig.~\ref{fig:1}c).}

\begin{figure}
 \includegraphics[width=0.5\textwidth]{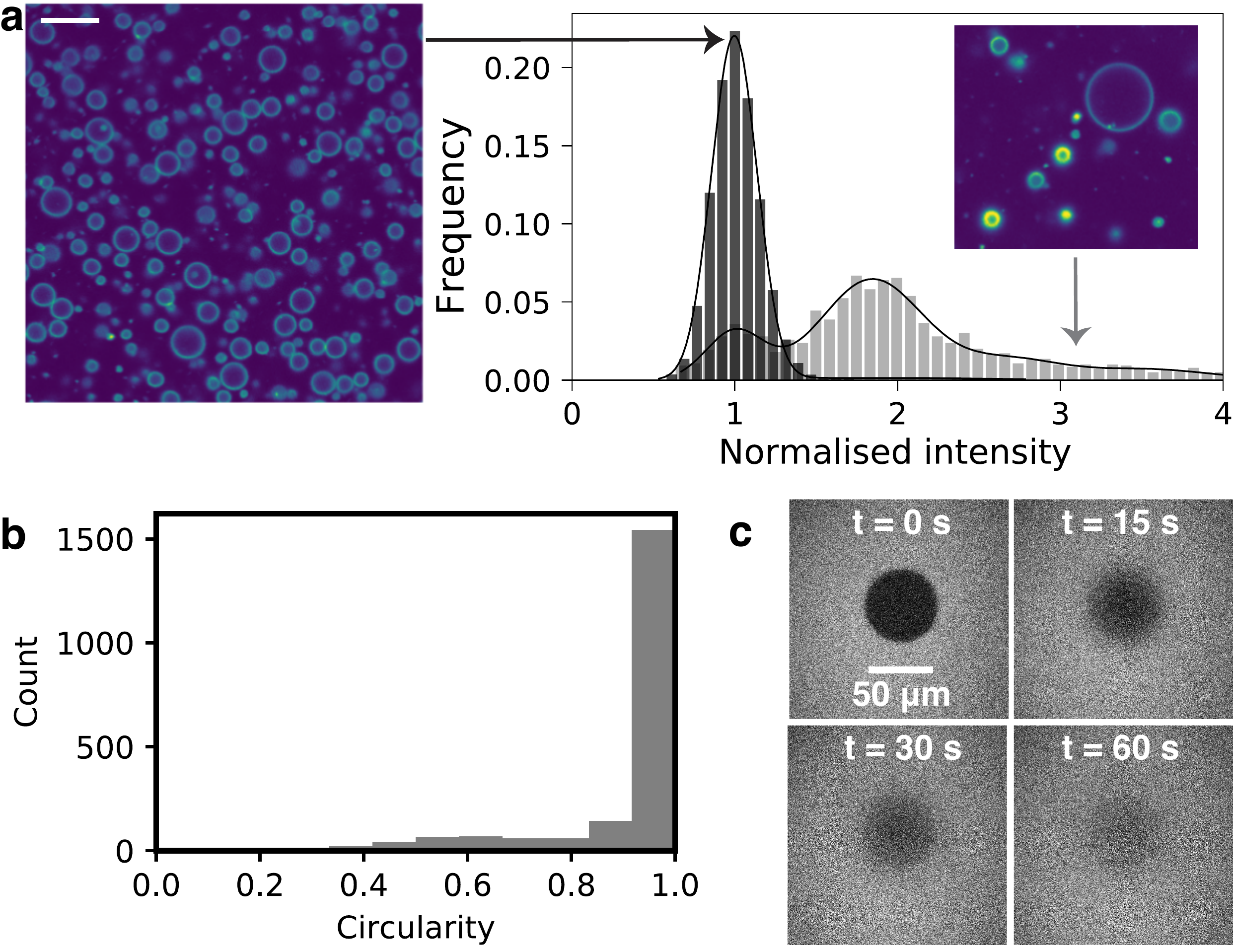}
\caption{Oleic acid can self-assemble into GUVs that look very \textcolor{black}{static}, but are in fact fluid. a. A histogram of mean membrane fluorescence intensities (see Methods) normalized against vesicles with the lowest mean membrane intensity per pixel for oligolamellar samples (grey, n = 1391) shows multiple peaks, whereas samples that appear uniform only have a single peak (black, n = 20347). Scale bar represents 10 $\mu$m. b. Oleic acid GUVs appear very spherical (see Methods). Circles have a circularity of 1. c. FRAP experiments revealed that oleic acid membranes have rapid lateral diffusion. A large bleached area (diameter 54 $\mu$m) was used to minimize the effect of diffusion during photobleaching (see also Fig.~S6).}
\label{fig:2}
\end{figure}

We used confocal microscopy to confirm that the cell-sized and apparently unilamellar vesicles observed with phase contrast microscopy were indeed unilamellar. We added the lipophilic dye, rhodamine B, to the vesicles after self-assembly, because the membrane fluorescence intensity is expected to reflect the amount of membrane material present. We found that an oligolamellar solution of vesicles showed clear differences in fluorescence intensity between vesicles (Fig.~\ref{fig:2}a). By contrast, the samples that appeared to have uniform membrane intensity under phase contrast microscopy exhibited a single peak in fluorescence intensity (Fig.~\ref{fig:2}a). These results are consistent with the latter samples being unilamellar~\cite{akashi_preparation_1996}.

Intriguingly, the GUVs did not show any \textcolor{black}{observable} membrane fluctuations and appeared extremely spherical (Fig.~\ref{fig:1}b, Fig.~\ref{fig:2}b, \textcolor{black}{Video~S1}). Although the melting temperature of oleic acid (T$_\mathrm{m} = 14^\circ$C) is below room temperature (T $\sim 21^\circ$C), the apparent \textcolor{black}{lack of membrane fluctuations} prompted us to verify that the membranes were not in a gel phase. FRAP (Fluorescence Recovery After Photobleaching) experiments on oleic acid GUVs resulted in apparently uniform bleaching across entire GUVs within 1 s. Thus to quantify the fluidity, we instead \textcolor{black}{performed FRAP experiments on} supported lipid bilayers with a large bleached area (Fig.~\ref{fig:2}c, \textcolor{black}{Fig.~S6)}, and found that the diffusion constant $D$ = 14.3 $\mu$m/s$^2$ $\pm$ 4.7 $\mu$m/s$^2$ (s.d., n = 7) is rapid compared to phospholipid bilayers. \textcolor{black}{The variation of dye diffusivity in these experiments may be owing to potential inhomogeneity of the supported lipid bilayers}. It is thus plausible that the bilayers appeared to \textcolor{black}{lack fluctuations} not because they were in a gel phase, but because the excess surface area required for surface undulations had somehow been eliminated.

\begin{figure}
 \includegraphics[width=0.5\textwidth]{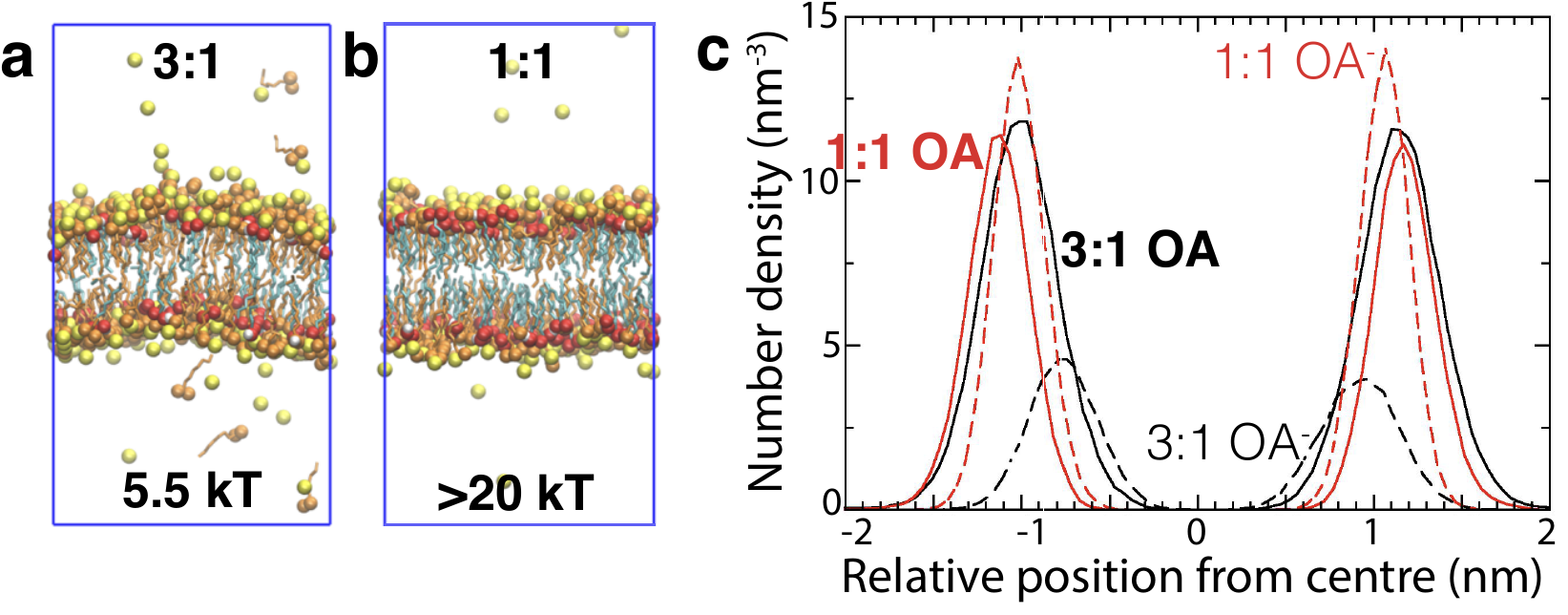}
\caption{Self-assembly outcomes are dictated by the protonation state of the membranes. a. Molecular dynamics simulations of an octanoic acid (OA) bilayer with three octanoates for every octanoic acid molecule (3:1) showed that such bilayers exhibited short-wavelength fluctuations, and have a bending modulus $K_c$ equal to 5.5 kT. b. Molecular dynamics simulations of an OA bilayer with one octanoate for every octanoic acid molecule (1:1) showed that such bilayers fluctuate less, and have a bending modulus $K_c$ greater than 20 kT. c. The profiles showing distributions of oxygen sites perpendicular to the bilayer are shown for both 3:1 and 1:1 systems. The 3:1 bilayer was thinner than the 1:1 bilayer.}
\label{fig:3}
\end{figure}

To understand the effect of protonation state on molecular packing in fatty acid bilayers, we carried out molecular dynamics simulations. We chose octanoic acid as the model lipid, because in our experiments the exact nature of the lipid tails did not seem to prevent the formation of GUVs. Linoleic acid, which has two degrees of unsaturation, could also self-assemble into GUVs, as could fatty acid mixtures containing both saturated and unsaturated species (Fig.~S7). We simulated octanoic acid bilayer membranes at 25\% protonation (3:1 sample, Fig.~\ref{fig:3}a) and 50\% (1:1 sample, Fig.~\ref{fig:3}b). For both 1:1 and 3:1 bilayers, the majority of protonated octanoic acid headgroups served as hydrogen bond donors to deprotonated octanoates, consistent with existing literature~\cite{dejanovic_surface_2011}. Much smaller fractions of octanoic acid in both cases were hydrogen-bonded to other octanoic acids or to solvent. While the great majority of molecules that formed the bilayer were hydrogen-bonded to others through their headgroups in the 1:1 bilayer, there were not enough donors for this to occur in the 3:1 bilayer. As a result, the 3:1 bilayer was thinner than the 1:1 bilayer (Fig.~\ref{fig:3}c), owing to increased repulsion between headgroups, which increased the area per headgroup. By analyzing the fluctuations in the membrane using the method of Brown \textit{et al.}~\cite{watson_determining_2012, levine_determination_2014}, we found that the 3:1 bilayer has a bending modulus 5.5 kT, similar to that of bilayer membranes made from surfactants with a comparable carbon-length to octanoic acid~\cite{jung_origins_2001}. By contrast, the 1:1 bilayer had a bending modulus $K_c$ greater than 20 kT, which is as stiff as phospholipid bilayers ($K_c \sim$ 20 kT)~\cite{boal_mechanics_2012}. Thus fatty acid bilayers are surprisingly stiff and moreover, their stiffness can change dramatically with protonation state. 

The dependence of membrane bending modulus on protonation state then led us to contemplate the role of bending modulus in the assembly of fatty acid vesicles. Having rigid membranes penalizes the formation of high-curvature, smaller vesicles~\cite{safran_statistical_2018, israelachvili_intermolecular_2011, boal_mechanics_2012}, and is thus consistent with our finding that we can generate \emph{giant} vesicles at pHs lower than the apparent p$K_a$, despite using an orbital shaker when agitation is known to decrease vesicle size even in phospholipid systems~\cite{reeves_formation_1969}. Furthermore, the vesicles remained cell-sized even when larger amounts of lipid were used. Instead of assembling into smaller vesicles, which would increase the translational entropy of the system, the lipids self-assembled into space-filling, almost jammed suspensions of giant vesicles that contained other giant vesicles within them (Fig.~S8, Videos S9, S10). These results confirm that bending energy, rather than entropy, determines the self-assembly pathway. 

There are several factors that could contribute to the remarkably high yield of \emph{unilamellar} vesicles at lower pH. Although fatty acid membranes at a pH lower than the apparent p$K_a$ may be as stiff as phospholipid bilayers, the individual fatty acids are not kinetically-trapped in a bilayer membrane as are phospholipids. It is well-known that the monomer desorption rate for fatty acid membranes depends on the lateral tension in the membranes~\cite{chen_emergence_2004, gruen_lateral_1982}. The is the basis for the observation that membranes that are under tension (for example, osmotically swollen vesicles) can grow at the expense of their relaxed counterparts (for example, osmotically relaxed vesicles). Tension on membranes can also be created by shear flow~\cite{kraus_fluid_1996}. We estimate~\cite{hubbe_adhesion_1981} that after the initial mixing stage, the maximum shear stress in our system when using an orbital shaker is approximately 0.3 N/m$^2$ (see Methods). Shear appears to be critical because the oleic acid GUV yield decreases to almost negligible (Fig.~S11) when orbital shaking at slower speeds, resulting in a maximum shear stress of 0.2 N/m$^2$ or less.

According to elastic theory, a piece of membrane under constant strain has a lateral membrane tension $\Sigma$ proportional to the bending modulus $K_c$ (see Methods)~\cite{deserno_fluid_2015, boal_mechanics_2012}. From this relation, we can conclude that any tension-driven effects will be more pronounced for vesicles with a higher bending modulus. Because the different bilayers of multilamellar vesicles will experience different amounts of shear~\cite{luo_deformation_2015}, if the pH is below the apparent p$K_a$ of the fatty acid, competition between adjacent layers may continue until only one membrane remains. For example, the more relaxed inner membranes may lose fatty acid molecules, which would tend to integrate into the more strained outer bilayers. Furthermore, the driving force for vesicle fusion increases at lower pH, owing to the increase in bending modulus, while the barrier to fusion presumably decreases owing to the decrease in surface charge density~\cite{safran_statistical_2018, israelachvili_intermolecular_2011}. Multiple fusion events between inner and outer membranes could lead to the formation of unilamellar vesicles. Indeed, foam-like intermediates were visible prior to GUV formation (Fig.~S12). Finally, after an increase in surface area from membrane remodelling, the requisite increase in volume to compensate for the excess surface area could come from the membrane momentarily rupturing, allowing the influx of suspended materials as large as colloidal particles. The excess surface area could also be lost by budding, as shown below. Consequently, a low pH not only favors the formation of \emph{giant} vesicles, but also of \emph{unilamellar} ones.

\begin{figure*}[t!]
 \includegraphics[width=\textwidth]{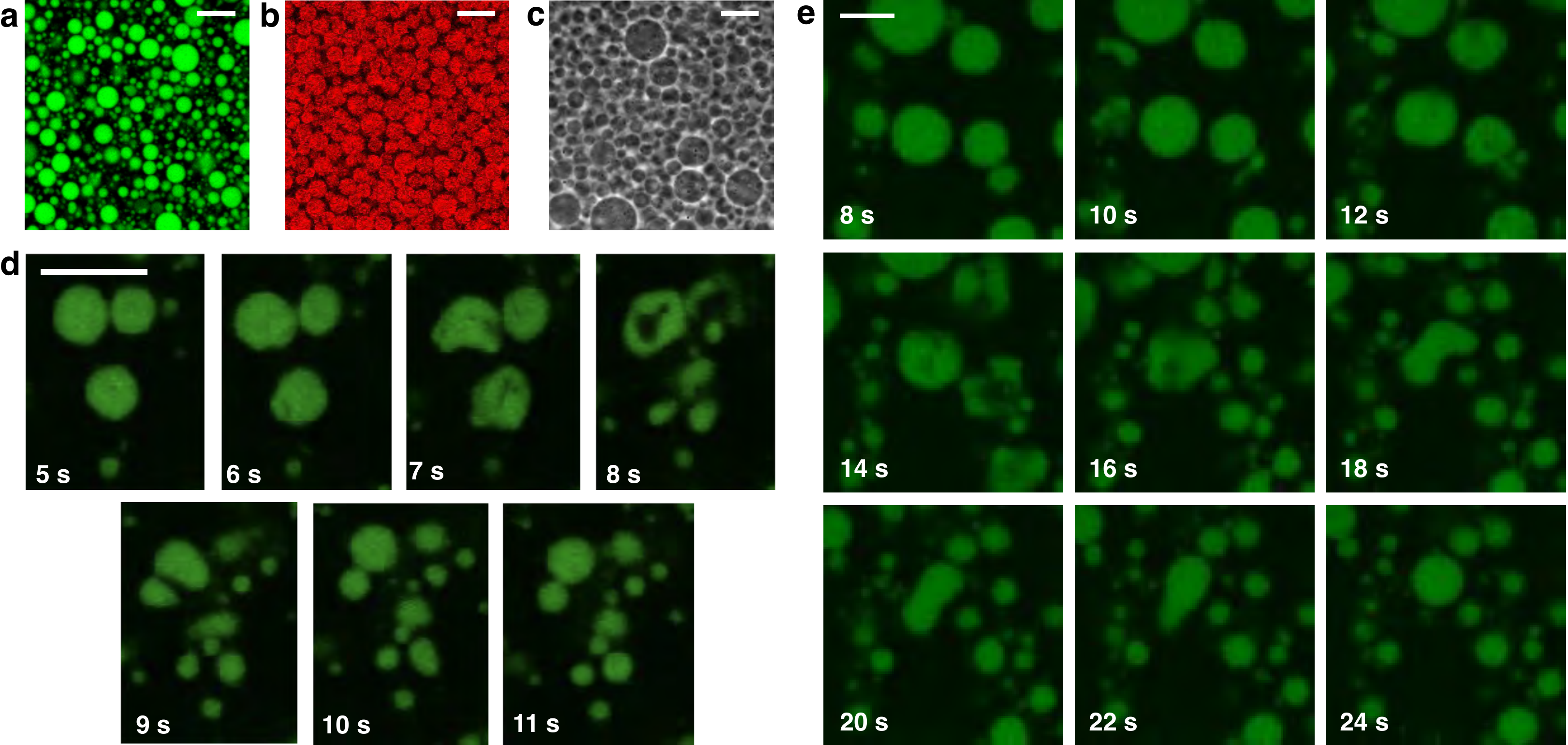}
\caption{Fatty acid GUVs can encapsulate a wide range of materials. Materials encapsulated include a. pyranine (1 mM), b. 30-nm-diameter Nile Red dyed latex nanoparticles (0.1\% w/v), and c. 400-nm-diameter polystyrene latex beads (0.1\% w/v; see also Video S13). d--e. Oleic acid vesicles can encapsulate RNA (the same RNA oligomer as in Fig.~\ref{fig:1}a) and divide upon exposure to oleate micelles pipetted nearby. The time shown represents the number of seconds after injecting oleate micelles. Scale bar represents 10 $\mu$m for all images.}
\label{fig:4}
\end{figure*}

We anticipate that fatty acid GUVs will be useful for a variety of applications in different fields because of their ease of self-assembly, the ready availability of fatty acids, and their versatility with respect to the encapsulation of contents. Fatty acid GUVs can be loaded with contents ranging from small molecules such as sugars (Fig.~\ref{fig:1}b), fluorescent dyes (Fig.~\ref{fig:4}a), and RNA oligomers (Fig.~\ref{fig:1}b), to nanoparticles (Fig.~\ref{fig:4}b) and colloidal particles (Fig.~\ref{fig:4}c, see also Video S13). After GUV assembly in the presence of each of these materials, the vesicles can be diluted into a new buffer, or washed to remove unencapsulated material (Fig.~\ref{fig:4}a--c). While the encapsulation of small solutes using bulk assembly techniques is routine~\cite{walde_giant_2010}, the encapsulation of colloidal particles is usually only possible using droplet-based techniques such as the emulsion transfer method or microfluidics~\cite{abkarian_continuous_2011}, \textcolor{black}{centrifugation during rehydration~\cite{natsume_shape_2010}, or layer-by-layer assembly methods to create polyelectrolyte capsules that either encapsulate or grow particles within~\cite{sukhorukov_physical_2004, radtchenko_inorganic_2002}. Despite the individual fatty acid molecules being very dynamic, being able to move between the membrane and solution, the vesicles themselves can withstand centrifuging at 500 g against a filter for at least 25 mins (Fig. S14), greatly simplifying the removal of unencapsulated material. Samples also appear unchanged over weeks (Videos S15 and S16)}. Our bulk-assembly technique is thus notable for its ability to encapsulate and retain contents across several orders of magnitude in size. 

One application of our findings is that we can now generate cell-sized, unilamellar model primitive cells (protocells) for dynamical studies using optical microscopy. Fatty acids have been prime candidates for constituents of primitive membranes for decades~\cite{hargreaves_liposomes_1978}, owing to their chemical simplicity and presence on meteorites, which confirms that their synthesis can be abiotic. To examine the potential for fatty acid GUVs to undergo physically-driven growth and division, we encapsulated fluorescently-tagged RNA inside of oleic acid vesicles to create simple protocells. We then added excess oleate micelles to the solution to enable an increase in the surface area to volume ratio as the added oleate became incorporated into pre-existing membranes. We found that rather than growing into tubular vesicles that require shear to divide, as previously reported for MLVs~\cite{zhu_coupled_2009}, our model protocells could grow and divide within a few seconds (Fig.~\ref{fig:4}d,e). The vesicles deformed by undergoing undulations, and higher frame-rate imaging (Video S17, S18) shows that the undulations represent the dynamic formation of lobes, which frequently appear to pinch off, forming smaller daughter vesicles. These intermediate structures are reminiscent of the recent theoretical findings of Ruiz-Herrero, Fai, and Mahadevan~\cite{ruiz-herrero_dynamics_2019}, whose models show that lobe-like structures can form in vesicles under certain growth conditions. Thus in our system, the intimate coupling of membrane growth with division can be attributed to the buildup of elastic energy when oleate is incorporated into the outer leaflet as follows. Because the membranes are made in the presence of salts with limited permeability, the volume is osmotically constrained at short timescales. Flip-flop across the membrane is known to relax the bending stress, and occurs with a rate of $\sim$0.5 s$^{-1}$~\cite{bruckner_flip-flop-induced_2009}. When the rate of membrane growth was potentially faster than the rate at which flip-flop could relieve the accumulated stress, the vesicles were able to undergo a dramatic fission event (Video S17, S18). When the rate of membrane growth was slower, fission did not occur (Video S19). \textcolor{black}{The ability of high curvature stresses to drive neck fission was recently reported for protein-laden membranes~\cite{steinkuhler_controlled_2020}}. We note that these results also demonstrate the strong propensity of fatty acid GUVs to remain spherical, consistent with large-scale membrane fluctuations being energetically unfavorable due to the high bending modulus~\cite{helfrich_steric_1978}.

We have shown that GUVs can be self-assembled in bulk by means distinct from the long-established picture of unilamellar vesicle self-asssembly being driven towards unilamellarity by Helfrich fluctuations and spontaneous curvature. Rather, fatty acid GUVs are driven towards a unilamellar architecture because the membrane is able to remodel despite being rigid. Whereas previous reports of unilamellar vesicles made by bulk self-assembly generated nanoscale vesicles, our vesicles are cell-sized, ideal for research using common biophysical tools such as optical microscopy. Because our results can be implemented with ease -- requiring minimal infrastructure, cost, chemicals, and skill -- we anticipate that these results will facilitate the study of a multitude of research questions in soft-matter science, origins of life, and biophysics.

\section{Methods}
\label{methods}
\subsection{Chemicals}
Oleic acid, palmitoleic acid, myristoleic acid, and decanoic acid were purchased from Nuchek-Prep (USA). Bicine (99\%), NaOH (solid, 99\%), HCl (37\%), pyranine, rhodamine B, sucrose, glucose, (3-aminopropyl)triethoxysilane (97\% Sigma-Aldrich) were purchased from Sigma-Aldrich (USA). Ammonium hydroxide (30\%), methanol (99.9\%), ethanol (99.5\%), isopropanol (99.9\%), chloroform (99.8\%), were obtained from Acros Organics (Fisher Scientific, USA). Other materials used include Hellmanex III (Hellma, Germany), TopFluor PC (Avanti Polar Lipids, USA), the oligonucleotide r(GGC UCG ACU GAU GAG GCG CG)-AF647 (IDT, USA), NOA 61 epoxy (Norland Products, Inc., USA) 30-nm-diameter Nile Red latex nanoparticles (Spherotech, Inc. USA), 400-nm-diameter latex size standard beads (Malvern, USA), All water used was Millipore (Millipore, USA). All chemicals were used as received. 

\subsection{Making vesicles}
100 mM fatty acid micelle stock solution were first prepared by adding 100 $\mu$mol of fatty acid to 1 mL of 125 mM NaOH in water. Bicine buffer stock solutions were adjusted to near the p$K_a$ of the fatty acid by NaOH addition. The vesicles were then prepared by adding micelle stock solution to a buffered solution, vortexing for 3 s, and then leaving the 0.5 or 1.7 mL microcentrifuge tube (Fisher Scientific, USA) or 20-mL glass scintillation vial (VWR) on an orbital shaker at 80 rpm (GeneMate BioExpress, USA and PSU-10i Grant Bio, UK) overnight. For all samples, 5 mM of fatty acid was used unless specified otherwise. For  myristoleic acid vesicle samples, 10 mM myristoleic acid was used. Any solute intended for encapsulation was included in the buffered solution (200 mM sucrose, 10 $\mu$M oligonucleotide, 1 mM pyranine, 0.1\% w/v nanoparticles, 0.1\% w/v colloidal particles). The resulting solution's pH was measured on a SevenCompact pH meter (Mettler Toledo, USA) or Orion pH meter (Thermofisher, USA).

\subsection{Washing vesicles}
To provide contrast between encapsulated contents and unencapsulated material during imaging, vesicles were prepared with 200 mM sucrose contained in the buffer and then washed as follows. After vesicle formation, the vesicles were diluted 1:10 into a 200 mM glucose buffer that was identical in composition and pH to the original buffer but lacked the lipids and encapsulated solute. After centrifugation at 2000 g for 30 s, the top 91\% of volume was removed by pipetting. The remaining solution was then agitated by flicking or pipetting up and down to resuspend the vesicles.

\subsection{Imaging vesicles}
Vesicles were imaged by phase contrast using a 1.4 NA 100$\times$ Plan Apo objective (Nikon, Japan) on a TE-2000 inverted microscope (Nikon, Japan). Images were captured with both an Ace CMOS (Basler AG, Germany) and Luca R (Andor Technology Ltd, UK) with similar results. Vesicles were also imaged by confocal microscopy on a A1R/Ti setup (Nikon, Japan). \textcolor{black}{Imaging glassware (22$\times$22 mm No. 1 and 24$\times$60 mm No. 1.5 glass coverslips (Fisher Scientific, USA)) was cleaned by soaking sequentially for at least one day each in 2 \% Hellamanex III, 70\% v/v isopropanol in water, then 2 M NaOH, with water rinses in between.}

\subsection{Membrane intensity quantification}
Unilamellar vesicles were prepared using 2.5 mM oleic acid in 200 mM sucrose and 50 mM bicine solution, pH 8.2, and left on an orbital shaker overnight at 80 rpm. Oligolamellar were prepared using 2.5 mM oleic acid in 200 mM sucrose and 200 mM bicine, pH 8.43, and left still for one week. They were then diluted 1:10 into an equi-osmolar glucose buffer in a Nunc Lab-Tek II 8-well chambered coverslip (Thermo Scientific, USA), then dyed by the addition of 40 $\mu$M rhodamine B. The vesicles were allowed to sediment for at least 30 mins, to create a monolayer of vesicles on the bottom for imaging.

The membrane intensity was quantified by a custom Python routine that detected circles by the cv2 Hough transform, and then measured the mean intensity of pixels in the circles. The intensities were then plotted as histograms, and fitted to Gaussian distributions. The intensities were then normalized to the mean of the lowest intensity Gaussian, and re-plotted as histograms.

\subsection{Circularity} Circularity of the GUV sample from Fig.~\ref{fig:4}a was quantified using ImageJ~\cite{schneider_nih_2012}. Circularity is defined as 4$\pi(A/P^2$), where $A$ is the area and $P$ is the perimeter.

\subsection{Shear stress} The maximum shear stress $\tau$, occuring adjacent to the bottom of a 20-mL scintillation vial on an orbital shaker is calculated using the relation from Hubbe~\cite{hubbe_adhesion_1981} $\tau = -\sqrt{\mu \rho a^2 \omega^3}$ where $\mu$ is the dynamic viscosity of the aqueous solution (0.89 mPas for water), $\rho$ is the density of the aqueous solution (997 kg/m$^3$ for water), $a$ is the amplitude of the rotational stroke (in our case, 13 mm for a 20-mL scintillation vial), and $\omega$ is the rotational speed in radians per second.

\subsection{Fluorescence Recovery After Photobleaching experiment}
Stock solutions of oleic acid and TopFluor-PC were prepared in chloroform, then deposited in a 20-mL glass scintillation vial (VWR, USA) and evaporated under a stream of nitrogen leaving 10 $\mu$mol of oleic acid and 20 nmol of TopFluor PC. 1 mL of 50 mM Na-bicine (pH 8.4) and 150 mM NaCl solution was then added to hydrate the lipid film, and left for 1 hr on a hotplate at 65$^\circ$C. The solution was cooled to room temperature, vortexed for 5 s, then withdrawn into a syringe (Hamilton Company, USA) and extruded 11 times through polycarbonate membranes with 100-nm-pores (Whatman) on a mini-extruder (Avanti Polar Lipids, USA). Meanwhile, 24$\times$60 mm No. 1.5 glass coverslips (Fisher Scientific, USA) were prepared by soaking sequentially for at least one day each in 2 \% Hellamanex III, 70\% v/v isopropanol in water, then 2 M NaOH, with water rinses in between. The silanisation solution of 30 mL ethanol, 1 mL (3-aminopropyl)triethoxysilane and 500 $\mu$L of 30\% ammonium hydroxide solution, was prepared fresh in a 50-mL tube (Falcon, USA). The cleaned glass coverslips were submerged into this solution for 2 mins before washing thoroughly with methanol and dried under nitrogen. The top lip of an open 500 $\mu$L microcentrifuge tube was cut off with a razor blade, then glued lip-down onto the cleaned coverslip with NOA 61 epoxy to create a cylindrical well. Curing time was 5 minutes at 365 nm and 4 W with a UVGL-15 Compact UV Lamp (UVP, USA), with the glass placed directly onto the lamp. Then 20 $\mu$L of the vesicle solution was deposited into the well, and allowed to spread onto the silanized glass for 60 minutes. Undeposited lipid was then washed three times by removing 90\% of the solution and pipetting in fresh buffer. The sample chamber was capped with a 18$\times$18 mm No. 1 glass coverslip (Fisher Scientific, USA) to minimize evaporation during imaging.

Imaging of the supported lipid bilayer was done on a A1R/Ti confocal microscope (Nikon, Japan). A FRAP routine on NIS Elements was used to acquire the FRAP image data. Lateral diffusion was rapid, so a large bleached region (54-$\mu$m-diameter) was used to minimize the effect of diffusion during photobleaching. Analysis was performed by following the method of Kang \textit{et al.}~\cite{kang_simplified_2012} \textcolor{black}{and the fits are shown in Fig.~S6}.

\subsection{Simulations}
Molecular dynamics simulations of 240 total molecules of octanoic acid/sodium octanoate at 1:1 and 1:3 ratios, with over 6000 TIP4P-2005 waters~\cite{abascal_general_2005} were performed using Gromacs~\cite{pronk_gromacs_2013}. Force field parameters for alkyl tails used the HH-Alkane model~\cite{ashbaugh_optimization_2011}, in which intramolecular and tail-tail interactions are taken from the TrAPPE united-atom model~\cite{martin_predicting_1997} and interactions between water oxygen and CH$_2$ and CH$_3$ sites adjusted to improve alkane hydration free energies.  Following Hess and van der Vegt~\cite{hess_cation_2009}, who used the Kirkwood-Buff formalism to validate ion-ion interactions for Na$^+$ with  carboxylate against experimental results, we use OPLS~\cite{jorgensen_development_1996} values for Lennard-Jones (LJ) parameters and partial charges of these ions and the carboxylate headgroup. OPLS parameters were also used for the carboxylic acid, with the partial charge on the COOH proton adjusted from +0.45 $e$ to +0.5 $e$.  The motivation for this change is that in our preliminary simulations on mixed-protonation state fatty acid bilayers, carboxylate and carboxylic acid headgroups were spatially separated into zones with different degrees of hydration (similarly to one previous simulation study~\cite{han_molecular_2013}) with very infrequent hydrogen bonding between headgroups. This result runs counter to experimental evidence for H-bond networks at the surface of fatty acid bilayers~\cite{dejanovic_surface_2011} and to the principle of PA/p$K_a$ matching~\cite{gilli_predicting_2009} (where PA stands for proton affinity), which would place RCOOH above water as a preferred hydrogen-bond donor towards RCOO$^-$.  The increase was sufficient to bring the gas-phase force field binding energies for trans- and cis-acetic acid to acetate up to 118 and 129 kJ/mol respectively, in line with calculated values from DFT (113-114 and 125 kJ/mol, for formic acid/formate~\cite{pan_characterization_1998, bako_carparrinello_2006}). 

A dodecanoic acid bilayer generated using the CHARMM-GUI membrane builder~\cite{jo_charmm-gui:_2008, wu_charmm-gui_2014} was used as a starting point for the C$_8$ bilayers. The initial distribution of protonated and deprotonated molecules was selected randomly; diffusion was observed to be fast on the simulation timescale for these FA bilayers (in contrast to phospholipids).  Simulations were performed with all bond lengths constrained to fixed distances using the LINCS algorithm~\cite{hess_lincs:_1997}. The pressure is maintained at 1 bar and surface tension at zero using semi-isotropic pressure coupling via the Berendsen barostat~\cite{berendsen_molecular_1984} with $\tau_P$ = 2 ps and compressibility of 4.5 $\times 10^{-5}$ bar$^{-1}$.  The Gromacs default (leap-frog) integrator with a 2 fs time step was used for integration of equations of motion. The Verlet~\cite{pall_flexible_2013} cutoff-scheme was applied for short-range non-bonded interactions with a cutoff of 1.4 nm. Particle-mesh Ewald summation~\cite{darden_particle_1993} was used to account for Coulomb interactions with a real space cutoff of 1.4 nm. The temperature was maintained at 300K by velocity rescaling thermostat~\cite{bussi_canonical_2007} with $\tau_T$ = 2 ps.  

\subsection{Bending modulus}
To estimate bending modulus $K_c$ from the simulation data we followed the method of Brown \textit{et al.}~\cite{watson_determining_2012, levine_determination_2014} by analyzing the fluctuations of fatty acid tail tilt vectors in Fourier space.  In the long wavelength/low-$q$ limit, the following relation should hold:

\begin{equation}
\langle \left |\hat{n}_q^{\parallel}\right |^2 \rangle = \frac{k_B T}{K_c q^2}
\end{equation}

where $\hat{n}_q^{\parallel}$ is the longitudinal Fourier mode along wavevector $q$ of the molecular tilt. To obtain this quantity, we took snapshots at 5 ps intervals starting at 15 ns and analysed the positions of C1 and C8 sites on all molecules, irrespective of protonation state.  The midplane of the bilayer was taken as the mean $z$ coordinate of all C1 sites. Molecules whose terminal methyl (C8) site was farther than 1 standard deviation from the midplane were excluded from the analysis; these included both molecules that escaped into the solvent and those with very nonstandard configurations, and constituted fewer than 5\% of all molecules at any time.  Molecules were assigned to a 12 $\times$12 grid based on the lateral position of their C1 site and a leaflet based on whether the $z$ coordinate of the C1 site was above or below the midplane.  A unit vector for each molecule $i$ was calculated as:

\begin{equation}
\mathbf{n}_i = \frac{(\mathbf{r}_{C8,i} - \mathbf{r}_{C1,i})}{|\mathbf{r}_{C8,i} - \mathbf{r}_{C1,i}|}
\label{eq:vec}
\end{equation}

A mean vector for each grid square was then calculated by subtracting the sum of the unit vectors in the lower leaflet from the sum of unit vectors in the upper leaflet in that grid square, and dividing by the total number of molecules in that square.  If a grid square was empty, the interpolation scheme described in Reference~\cite{levine_determination_2014} was used to assign its mean vector.  A two-dimensional fast Fourier transform of the $x$- and $y$-coordinates of the grid square vectors was taken for each frame, and the longitudinal component of this tilt vector fluctuations are found by projecting the $x$- and $y$- components along the $q$-vector.  The average of the resulting square amplitudes, taken over all frames and over wavevectors with the same magnitude, is then taken. Rearranging Eq.~\ref{eq:vec} yields

\begin{equation}
\frac{K_c}{k_B T} = q^{-2} \langle \left | \hat{n}_q^{\parallel}\right |^2 \rangle^{-1}
\end{equation}

\subsection{Elastic theory}
A piece of bilayer membrane that occupies an area $A$ under stress (and an area $A_0$ under no external stress) has a dimensionless strain $u = A/A_0 - 1$. The lateral tension of the membrane $\Sigma$ is related to the membrane's stretching modulus $K_\mathrm{stretch}$ by $\Sigma = K_\mathrm{stretch} u$. Assuming the bilayers are two uncoupled elastic sheets~\cite{deserno_fluid_2015, boal_mechanics_2012}, the stretching modulus is related to the bending modulus $K_c$ by $K_\mathrm{stretch} \propto K_c/h^2$ where $h$ is the bilayer thickness. This leads us to the relation between membrane tension and bending modulus $\Sigma \propto {K_c u}/{h^2}$.

\begin{acknowledgments}
A.W. would like to acknowledge the support of the NASA Postdoctoral Program Fellowship in Astrobiology and the UNSW Sydney Scientia Fellowship. J.T.K. acknowledges computational resources of the Extreme Science and Engineering Discovery Environment (XSEDE) Comet cluster at the San Diego Supercomputer Center, which is supported by the National Science Foundation Grant No. ACI-1548562, Allocation No. TG-MCB110144. J.W.S. is an Investigator of the Howard Hughes Medical Institute.  This work was supported in part by a grant (290363) from the Simons Foundation to J.W.S.
\end{acknowledgments}

\subsection*{Author Contributions}
All authors contributed to the writing of the manuscript and interpretation of the data. J.T.K. conceived of and did the molecular dynamics simulations and subsequent analyses. J.W.S. conceived of the vesicle division experiments. A.W. conceived of, did, and analysed the experiments.

\subsection*{Competing Interests Statement}
The authors declare no competing interests.

\subsection*{Data availability statement}
 The data supporting the findings of this study are
available within the paper and its supplementary information files. Additional data that support the findings of this study are available from the corresponding authors upon reasonable request.

\subsection*{Code availability}
Code used to analyse the membrane intensity, as well as the diffusion constant from FRAP data, are available from the corresponding authors upon reasonable request.

\subsection*{Materials and Correspondence}
\textit{Supplementary Information} is available for this paper.
Correspondence and requests for materials should be addressed to \url{anna.wang@unsw.edu.au}.

\bibliography{manuscript}
\end{document}